\documentclass[10pt,twocolumn,a4paper]{article}
\usepackage[a4paper,margin=25.4mm,columnsep=6mm]{geometry}
\usepackage{iftex}
\ifPDFTeX
  \usepackage[T1]{fontenc}
  \usepackage[utf8]{inputenc}
  \DeclareUnicodeCharacter{2212}{\ensuremath{-}}
\else
  \usepackage{fontspec}
\fi
\usepackage{lmodern}
\usepackage{amsmath,amssymb,graphicx,booktabs,tabularx,array}
\usepackage[font=small,labelfont=bf,justification=raggedright,singlelinecheck=false]{caption}
\usepackage{microtype,xurl,ragged2e}
\usepackage{xcolor}
\definecolor{linkblue}{RGB}{22,70,108}
\usepackage[colorlinks=true,allcolors=linkblue,pdfauthor={Frank Li},pdftitle={Shared KV Caching for Replicated 27B Inference: Correctness Failures and Performance Boundaries}]{hyperref}

\newcommand{\paperCode}[1]{\mbox{\texttt{\detokenize{#1}}}}
\makeatletter
\renewcommand{\section}{\@startsection{section}{1}{\z@}{-3.5ex plus -1ex minus -.2ex}{2.3ex plus .2ex}{\normalfont\Large\bfseries\raggedright\hyphenpenalty=10000\exhyphenpenalty=10000}}
\renewcommand{\subsection}{\@startsection{subsection}{2}{\z@}{-3.25ex plus -1ex minus -.2ex}{1.5ex plus .2ex}{\normalfont\large\bfseries\raggedright\hyphenpenalty=10000\exhyphenpenalty=10000}}
\makeatother
\title{Shared KV Caching for Replicated 27B Inference: Correctness Failures and Performance Boundaries}
\author{Frank Li\\UNSW Sydney\\\href{mailto:research@n1a.net}{\nolinkurl{research@n1a.net}}}
\date{}
\begin{document}
\raggedbottom
\twocolumn[
\begin{center}
{\LARGE\bfseries Shared KV Caching for Replicated 27B Inference: Correctness Failures and Performance Boundaries\par}
\vspace{10pt}
{\large Frank Li\par}
\vspace{2pt}
{\normalsize UNSW Sydney\par}
\vspace{1pt}
{\normalsize \href{mailto:research@n1a.net}{\nolinkurl{research@n1a.net}}\par}
\end{center}
\begin{minipage}{\textwidth}
\begin{abstract}
Shared host-memory caching can avoid repeated prefill when a request moves between inference replicas. Its usefulness depends on both correct state transfer and lost prefix locality. We study two single-GPU 27B vLLM replicas sharing a 256 GiB LMCache pool. After adopting an existing packed-page patch, we isolate a raw-pointer fallback that omits the dependency on the current CUDA stream. Controlled byte tests fail under an imposed delay and pass when the dependency is restored; the existing mixed allocator provides a working deployment path. Full-pool allocation checks and service regression complete the validation. A four-block OFF--ON--ON--OFF comparison contains 768 measured requests within two block pairs. Median cross-replica time to first content token falls from 31.715 to 0.605 seconds at 128k input and from 92.047 to 0.790 seconds at 256k. Six-turn synthetic sessions alternating replicas improve by approximately 35\% and 45\% at initial contexts of 32k and 128k, while fixed placement shows little benefit. This engineering case study identifies practical validation steps and the locality conditions in which shared caching pays off.
\end{abstract}
\vspace{6pt}
\end{minipage}
]
\RaggedRight
\section{Introduction}\label{introduction}

A multi-turn inference request commonly contains substantial context already processed in an earlier turn. A replica can reuse its local prefix state, but a different replica may need to recompute that prefix. This creates a concrete deployment choice: preserve request locality through placement, provide a shared state-reuse path, or combine both.

LMCache provides mechanisms for moving and reusing KV caches across requests and inference engines.\cite{ref1} We examine two deployment questions: how to validate the resulting state-transfer path, and how its performance compares with preserving a conversation on its original GPU.

Two issues motivate the study. First, a cache can appear operational while returning incorrect state: an HTTP success or a hit counter is not an integrity check. Second, an impressive cold-versus-warm comparison can obscure a competitive alternative---retaining the conversation on its original GPU. A useful evaluation must therefore establish correctness before measuring performance and must retain local-hit baselines.

The study contributes a controlled diagnosis linking copy ordering to an allocator-selected fallback, validation across bytes, runtime helpers, full-size allocation and service responses, and paired measurements separating cold computation, local reuse and cross-replica reuse. Together, these results provide a concrete basis for configuring and evaluating shared caching in a replicated service.

\section{System and Related Work}\label{system-and-related-work}

\begin{figure*}[!t]
\centering
\includegraphics[width=1.0\textwidth]{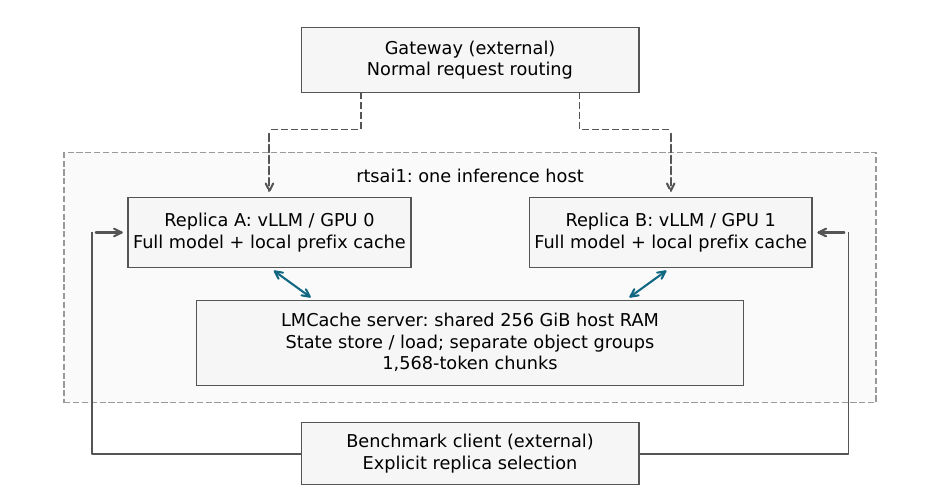}
\caption{Two complete inference replicas share a local RAM cache. Dashed arrows show normal gateway request routing; the controlled benchmark selects replicas directly. State store/load traffic remains on the inference host. The diagram describes logical paths, not measured transport bandwidth.}
\label{fig:1}
\end{figure*}

Figure~\ref{fig:1} shows the gateway and two complete vLLM replicas on one inference host. Each replica occupies one GPU and performs both prefill and decode. In the shared configuration, both connect to one host-memory LMCache service. Each retains its own GPU prefix cache. This is not prefill/decode disaggregation, and the gateway does not serve as the KV-data storage tier.

LMCache provides the cache-transfer and reuse mechanisms; our study tests their execution in a fixed deployment.\cite{ref1} Marconi addresses hybrid-state reuse constraints through admission and eviction policies.\cite{ref2} We take the deployed runtime's aligned recovery boundaries as given and test the transfer path used to preserve that state. Production characterization examines reuse patterns across real traffic and derives a workload-aware eviction policy.\cite{ref3} Our synthetic workloads control prefix reuse and replica placement to isolate locality-dependent effects. The comparison is with local GPU caching on the same deployment, holding the existing policies fixed.

\subsection{Evaluated configuration}\label{evaluated-configuration}

Table~\ref{tab:1} summarizes the evaluated configuration.

\begin{table*}[!t]
\centering\small\setlength{\tabcolsep}{5pt}
\caption{Evaluated configuration; current model files were verified on September 13. ``FP8 KV'' is a serving setting distinct from weight precision and recurrent-state representation.}
\label{tab:1}
\begin{tabularx}{\textwidth}{>{\raggedright\arraybackslash}p{.20\textwidth}>{\raggedright\arraybackslash}X}
\toprule
\textbf{Component} & \textbf{Recorded configuration} \\
\midrule
GPUs under test & Two NVIDIA RTX PRO 6000 Blackwell Server Edition GPUs; 97,887 MiB reported per device \\
Host CPU & Two AMD EPYC 9455 48-core processors; 192 logical CPUs; eight NUMA nodes \\
Driver & 610.57.04 \\
Model identifier & Local alias \paperCode{qwen3.8-27b-nvfp4}; export directory \paperCode{Qwen3.8-27B-NVFP4-v1}; current weights and metadata match fixed public \paperCode{kyaky/Qwen3.8-27B-NVFP4} \cite{ref9} \\
Text architecture & 64 layers: 48 Gated DeltaNet and 16 full attention; hidden size 5,120; full-attention head dimension 256 and four KV heads \cite{ref9,ref10} \\
Weight precision & Compressed-tensors export: FP8 large attention/GDN projections, NVFP4 MLP weights in layers 0--62, and retained higher-precision components including BF16 final-layer MLP and lm\_head \\
Tokenizer & Exported \paperCode{Qwen2Tokenizer} configuration, tokenizer JSON, and chat template; identifiers recorded in the post-experiment audit \\
Runtime & vLLM 0.28.0; LMCache 0.5.4; PyTorch 2.13.0+cu129 \\
Replica settings & Tensor parallelism 1; FP8 KV; HND layout; Mamba alignment; prefix caching enabled \\
Scheduling limits & 8,192 batched tokens; 64 sequences; maximum model length 1,010,000 \\
Reported GPU-cache capacity & 1,262 blocks and 1,957,941 cacheable tokens per replica, unchanged across four blocks \\
Shared tier, ON only & 256 GiB RAM; 1,568-token chunks; four kernel groups arranged into two object groups; LRU \\
Native RAM offload & Disabled in both arms \\
\bottomrule
\end{tabularx}
\end{table*}

Runtime GPU capacity is reproduced as reported; the paper does not infer usable capacity by multiplying nominal block parameters. The host contains eight GPUs, but only GPUs 0 and 1 are the experimental replicas. Other host activity is addressed in Section 6.

On September 13, a complete SHA-256 audit matched the 23,285,140,684-byte weight file and seven auxiliary files to public checkpoint revision \paperCode{6de592a7}\allowbreak{}\paperCode{a4a5618b}\allowbreak{}\paperCode{87c952ab}\allowbreak{}\paperCode{92928ee1}\allowbreak{}\paperCode{755bd9c6}.\cite{ref9} File metadata remained stable during reading. The publisher and retained local launcher identify the base as \paperCode{Qwen/Qwen3.8-27B}, whose configuration also uses the implementation class \paperCode{Qwen3_5ForConditionalGeneration}.\cite{ref10} This verifies current disk files; the September 12 performance run did not freeze full-weight hashes or an immutable export/calibration record. The dated {full-weight audit} supplements the earlier {metadata audit}.

The export mixes FP8 large attention/GDN projections (128×128 weight blocks and dynamic group-128 FP8 inputs) with group-16 NVFP4 MLP weights in layers 0--62. Those MLP groups have no input-activation quantization in the configuration; final-layer MLP, lm\_head, and other excluded components retain higher precision. ``NVFP4'' in the checkpoint name should not be read as uniform model-wide W4A4. Only text requests were evaluated.

\subsection{Shared hybrid state and recovery boundaries}\label{shared-hybrid-state-and-recovery-boundaries}

Here, ``KV cache'' includes hybrid decoding state recoverable at a common prefix boundary. Full attention contributes preceding K/V pages; Gated DeltaNet contributes boundary snapshots of convolution and recurrent state. vLLM manages these through its Mamba cache interface; that interface name does not change the model's architecture.\cite{ref11} The evaluated patch represents an entire recurrent page, including convolution state, recurrent state, and padding, as opaque addressable bytes. Its attention-shaped view does not imply per-token K/V or numerical INT8 quantization.\cite{ref4}

LMCache kernel groups collect layers with compatible transfer layouts; object groups collect kernel groups for storage under shared window requirements.\cite{ref11} Historical registration contains four kernel groups of 16 layers. Kernel groups 0--2 have a 1,568-token window and map to object group 0; full-prefix attention kernel group 3 maps to object group 1. Registration indices differ from original decoder layer order. Under \paperCode{align}, recovery requires both the needed attention-prefix blocks and the corresponding recurrent snapshot. A hit on the attention kernel group alone does not establish recovery of the entire hybrid state. The reported complete shareable prefix uses this aligned interpretation; byte equivalence across all kernel groups was not directly tested.

\section{Correctness Failures and Repairs}\label{correctness-failures-and-repairs}

\subsection{Packed-page compatibility}\label{packed-page-compatibility}

An initial cross-replica response returned corrupted text despite HTTP 200 and 25,088 externally hit tokens. The compatibility investigation found that a logical full-attention block comprised multiple packed kernel pages. The original handling recognized rank-five attention tensors but omitted the relevant rank-four packed representation, resulting in an incorrect view of the bytes to transfer.

We adopted the existing upstream packed-subpage patch identified by commit \paperCode{f180b9ff}\allowbreak{}\paperCode{ce7df45c}\allowbreak{}\paperCode{e3037011}\allowbreak{}\paperCode{d95a22db}\allowbreak{}\paperCode{947fefcb}, without additional local source modifications to that patch.\cite{ref4} In the recorded 27B case, a 1,568-token logical block spans 49 pages of 32 tokens. The patch supplies a logical-block view and validates layout constraints. HND was required for the evaluated unmodified patch and vLLM/FlashInfer combination: the NHD startup attempt failed with a kernel-page-count divisibility error, reflecting a mismatch between the patch's interpretation and the backend's logical shape/strides. This is a configuration-specific requirement, not a general requirement of LMCache. The {archived failure} and patch explanation retain the evidence. This repair addresses representation and is distinct from the copy-ordering issue below.

\subsection{A reproducible stream-ordering fault}\label{a-reproducible-stream-ordering-fault}

After the representation repair and a separate workaround registering the entire host pool together, a 128k service failure remained. Identical token input produced \paperCode{alpha=938639} when cold or locally reused on A. B externally hit 127,008 tokens but returned \paperCode{938624}; B-local reuse retained the error. A new cache salt forced zero-hit cold computation on B without altering the input and restored the correct answer. These {service controls} directed the investigation toward shared-state transfer and reuse after the registration-boundary fault had been controlled.

The evaluated image lacks the LMCache CUDA-ops extension and follows a PyTorch fallback. Its raw-pointer \paperCode{lmcache_memcpy_async} branch calls \paperCode{cudaMemcpy(}\allowbreak{}\paperCode{...,}\allowbreak{}\paperCode{ cudaMemcpyDefault)} without a current-stream argument; \paperCode{cudaMemcpyDefault} is the copy-kind enum, not the function name.\cite{ref5} The evaluated PyTorch stream pool uses \paperCode{cudaStreamNonBlocking}.\cite{ref6} Thus the non-default stream carrying the reproduced gather/scatter dependencies is excluded from legacy default-stream implicit synchronization; host-side return from the copy does not establish the missing producer/consumer dependency.\cite{ref7}

A 1 MiB device-to-host probe schedules a delayed write of byte value 37, then calls the fallback. Host memory receives the old value 0. A host-to-device probe schedules a read of the old value 0, then overwrites the temporary buffer with 53 through the fallback; the consumer reads 53 instead. Each probe reports 1,048,576 mismatching bytes. These are controlled integer-byte failures, not differences attributable to floating-point arithmetic or language-model sampling.

Supplementary controls on September 13 hold the pinned host and GPU buffers fixed while varying only the ordering mechanism. Two independent processes on an idle third GPU of the same model test 4 KiB, 1 MiB and 16 MiB objects in both directions, using either zero or \(10^8\) CUDA sleep cycles, with ten trials per cell. The raw fallback fails all 120 delayed checks and passes all 120 undelayed checks. Three controls---waiting for the preceding stream work before the same fallback, using \paperCode{cudaMemcpyAsync} on the current stream, and same-stream tensor copying---pass all 720 checks. Restoring ordering while retaining the buffers removes the reproduced fault. The imposed delay exposes the dependency; these counts are not production error rates.

The original helper-level probe exercises actual LMCache allocators and production GPU helpers. For a single 1 MiB uint8 object, the lazy path fails ten checks on GPU 0; the mixed path passes ten on GPU 0 and ten on GPU 1. Each set contains five checks per direction. This connects the byte mechanism to the branch used by the deployed configuration.

\begin{figure*}[!t]
\centering
\includegraphics[width=1.0\textwidth]{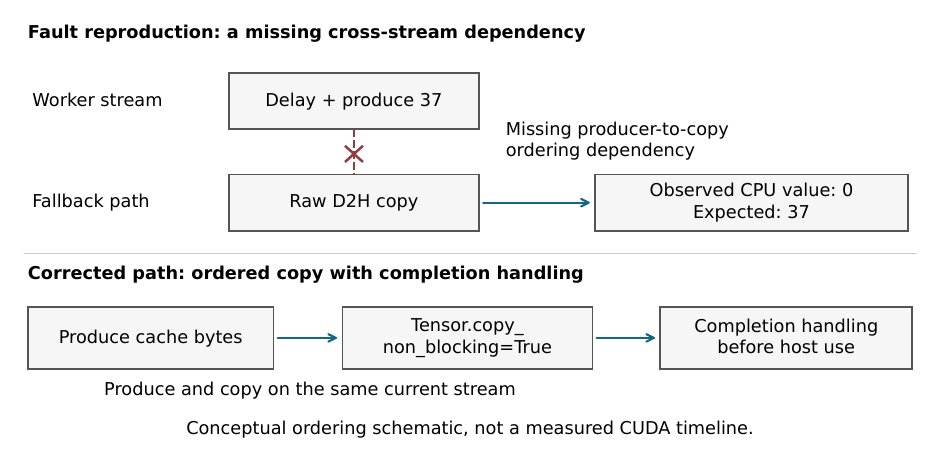}
\caption{The upper panel illustrates the missing dependency reproduced by the device-to-host probe. The lower panel shows ordered tensor operations on the same current stream. This is a conceptual dependency diagram, not a profiler trace or a time-scaled reconstruction of the original failed service request. Host-side completion handling remains necessary before host consumption.}
\label{fig:2}
\end{figure*}

Figure~\ref{fig:2} summarizes the ordering mechanism. Explicitly setting \paperCode{--no-l1-use-lazy} selects the existing mixed allocator and its tensor-copy path. Removing the positive lazy flag alone is insufficient because the evaluated CLI defaults to lazy allocation.\cite{ref5} The workaround bypasses the raw fallback, while completion handling still precedes host consumption.

Related upstream reports discuss staging-buffer ordering after tensor conversion and premature reuse of asynchronous-copy buffers.\cite{ref13,ref14} Those reports concern different transfer paths. Here, the controlled diagnosis isolates the lazy allocator's raw-pointer fallback. Current-stream copying already exists in LMCache's native implementation; the explicit-stream control checks that existing contract.\cite{ref5}

\subsection{Full-pool pinned allocation}\label{full-pool-pinned-allocation}

A small mixed-allocator probe successfully pins memory, but the first 256 GiB service allocation falls back to pageable memory. The allocator requests 256 GiB plus 4096 bytes for alignment. The evaluated PyTorch host allocator's power-of-two rounding can increase this request to 512 GiB.\cite{ref8} The service log preserves a generic allocation failure rather than the complete underlying exception, so the evidence does not establish rounding as the sole possible cause.

Setting \paperCode{PYTORCH_}\allowbreak{}\paperCode{CUDA_}\allowbreak{}\paperCode{ALLOC_}\allowbreak{}\paperCode{CONF=}\allowbreak{}\paperCode{pinned_}\allowbreak{}\paperCode{max_}\allowbreak{}\paperCode{round_}\allowbreak{}\paperCode{threshold_}\allowbreak{}\paperCode{mb:}\allowbreak{}\paperCode{128} limits large-allocation rounding without reducing the LMCache pool; the fixed allocator checks this threshold before rounding.\cite{ref8} In an isolated process using the same image, the full 274,877,911,040-byte allocation reports \paperCode{is_pinned()=true} and completes in 38.973 seconds. The subsequently configured service initializes its pool in approximately 42 seconds without a pinned-allocation fallback warning.

Pinning and copy ordering are separate properties. The mixed/pageable stage already passes service regression; pinning addresses the actual memory path used for the final performance evaluation. The manuscript does not treat timings from the intermediate pageable stage as part of the final ON/OFF experiment.

\subsection{Layered validation}\label{layered-validation}

Table~\ref{tab:2} summarizes the validation layers and their scope.

\begin{table*}[!t]
\centering\small\setlength{\tabcolsep}{5pt}
\caption{Copy controls include the September 13 supplement; 120 undelayed raw checks also pass. September 12 service stages total 264 non-warmup requests, with 12 warmups recorded separately.}
\label{tab:2}
\begin{tabularx}{\textwidth}{>{\raggedright\arraybackslash}p{.19\textwidth}>{\raggedright\arraybackslash}p{.43\textwidth}>{\raggedright\arraybackslash}X}
\toprule
\textbf{Layer} & \textbf{Recorded observation} & \textbf{Interpretation} \\
\midrule
Packed representation & Upstream compatibility tests pass after the adopted patch & Covers the tested logical-page layout \\
Copy-ordering controls & Delayed raw: 120/120 failures; ordered controls: 720/720 passes & Same buffers; three ways to restore ordering \\
Actual allocator/helper & Lazy: 10/10 failures; mixed: 20/20 passes across two GPUs & Tests a single object through the selected runtime path \\
Intermediate service stage & 207 non-warmup exact-answer requests pass & Mixed/pageable regression \\
Final pinned service stage & 57 non-warmup exact-answer requests pass & Final configuration regression \\
Protocol checks & 11 checks across the recorded stages pass & Tested tool-call and streaming behavior \\
\bottomrule
\end{tabularx}
\end{table*}

The service checks cover the original failing long-context fixture, directional reuse, exact retrieval of inserted values, and interleaved multi-turn sessions. They connect the runtime-path change to successful service regression.

\section{Experimental Methodology}\label{experimental-methodology}

\subsection{Comparison and isolation}\label{comparison-and-isolation}

The final screening uses four independently restarted blocks in the order OFF--ON--ON--OFF, forming two ON/OFF block pairs. OFF retains GPU prefix caching but removes the LMCache connector/dependency and stops the cache service. ON uses the repaired pinned configuration. Native RAM offload is disabled in both; OFF is not the historical native-offload deployment.

The evaluated model settings and reported GPU-cache capacity are held constant. Normal requests to both upstreams are placed in maintenance for measurement, and a guard records no need to reapply isolation. For single-request metric windows, prefix-query increments match the reported input-token count. Concurrent counters are interpreted only at window level.

Each block contains 192 measured requests and six warmups: 768 measured requests and 24 warmups in total. All 384 ON/OFF request-phase pairs match both payload SHA-256 and target GPU. The 624 retained fixture-file paths represent 312 unique payload digests; repeated phases reuse inputs, so the pair count is not a count of independent texts or deployments. Of the measured requests, 736 pass exact-value response assertions; 32 forced-length generation requests satisfy the specified 512/2048 output-token count, without semantic-quality scoring.

\subsection{Workloads}\label{workloads}

Microbenchmarks use nominal input sizes of 8k, 32k, 128k, and 256k tokens; actual usage counts are retained. Synthetic file-like text contains three values near its beginning, middle, and end, requested as short JSON. Requests use temperature 0, seed 314159, thinking disabled, and a 64-token normal output limit. Each chain executes cold computation, two local reuses, cross-replica reuse, and receiver-local reuse. There are four paired chains per size across the two block pairs. All cold requests have zero local and external hits. At explicit cross-replica steps, the receiving GPU has zero local hits; ON hits the complete shareable aligned prefix and OFF has zero external hits.

Session experiments start at 32k or 128k, append 1,055 tokens per later turn in the recorded inputs, and run for six turns. Responses contain short JSON with exact values. Canonical pregenerated history keeps ON/OFF inputs identical by inserting the expected JSON as the preceding assistant answer. This means later-turn success need not require retrieving the values from the original long prefix. Fixed and alternating placement use independent namespaces at the same nominal length target: actual initial counts are 31,991/31,993 and 128,000/127,980 tokens, respectively. They are not identical-payload pairs with each other. There are four paired sessions per size and routing condition.

Generation experiments use 8k/128k inputs, cold/hot prefixes, and \paperCode{ignore_eos=true} to force 512/2048-token outputs. This is a decode stress workload, not a validated normal long-answer task. Each cell has two paired observations, with GPU assignment swapped in the second block pair. Concurrent screening uses the same canonical-history construction with four or eight 32k-initial sessions of four turns each, under fixed and alternating placement. Each window contains only 16 or 32 requests, with two paired windows per condition.

\subsection{Metrics and statistical units}\label{metrics-and-statistical-units}

TTFT is measured at the client from request initiation to the first received content token. It includes request handling and transport and is not a direct CUDA prefill or transfer timer. Decode rate is

\begin{samepage}

\[
R_{\mathrm{decode}} = \frac{N_{\mathrm{out}}-1}{t_{\mathrm{last}}-t_{\mathrm{first}}}.
\]

Here \(N_{\mathrm{out}}\) is the output-token count, and \(t_{\mathrm{first}}\) and \(t_{\mathrm{last}}\) are the client-observed times of the first and last content tokens. SSE batching can affect this rate.

\end{samepage}

Session wall time includes request handling, client parsing, and recording, but excludes fixture-tokenization preparation and window-metric collection. No artificial tool delay is inserted. It is not a coding-agent task-completion metric.

We report descriptive medians, individual observations, and the mean of \(100\times(\mathrm{ON}/\mathrm{OFF}-1)\) over matched units. Each cell contains four chain/session pairs or two generation/window pairs, nested within two independently restarted block pairs. Requests within a block share execution conditions. We therefore report the observed effects without confidence intervals or significance claims; request counts do not increase the number of independent block pairs.

\subsection{Excluded trial}\label{excluded-trial}

An initial attempt reused a namespace across input sizes, causing a purportedly cold 128k request to hit 10,976 local tokens. The assertion stopped the run. The complete affected block is archived and excluded; individual unfavorable measurements were not selectively removed. Namespaces were corrected to include size, caches were cleared through restarts, and the entire four-block screening was rerun.

\section{Results}\label{results}

\subsection{Locality determines the observed benefit}\label{locality-determines-the-observed-benefit}

\begin{figure*}[!t]
\centering
\includegraphics[width=1.0\textwidth]{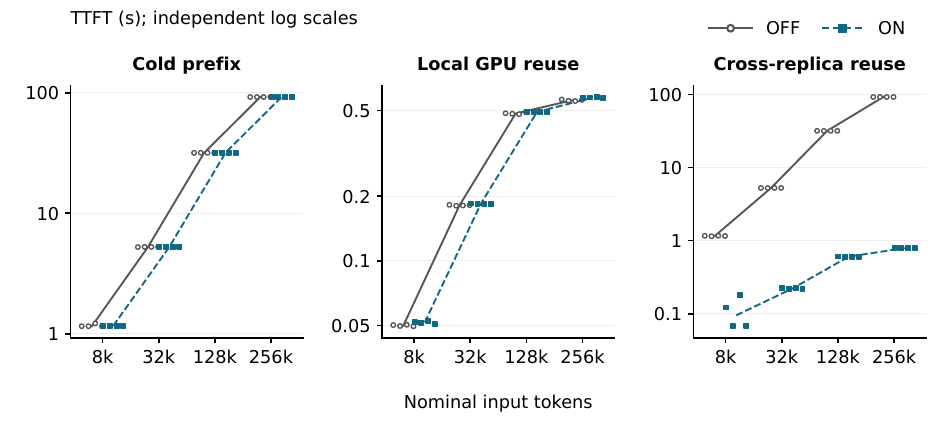}
\caption{Cold, first local-reuse, and cross-replica TTFT. Points show four observations per configuration and input size; lines show medians. Horizontal offsets separate configurations and observations and have no quantitative meaning. All y axes are logarithmic and independently scaled. Observations are nested within two paired blocks, not four independent machine allocations.}
\label{fig:3}
\end{figure*}

Figure~\ref{fig:3} and Table~\ref{tab:3} compare TTFT across locality conditions.

\begin{table*}[!t]
\centering\small\setlength{\tabcolsep}{2pt}
\caption{Medians from four paired chains per nominal size. ``Local'' is the first same-replica reuse. Each chain contributes measurements to all three locality conditions.}
\label{tab:3}
\begin{tabular*}{\textwidth}{@{\extracolsep{\fill}}lrrrrrr@{}}
\toprule
\textbf{Nominal input} & \textbf{OFF cold (s)} & \textbf{ON cold (s)} & \textbf{OFF local (s)} & \textbf{ON local (s)} & \textbf{OFF cross (s)} & \textbf{ON cross (s)} \\
\midrule
8k & 1.157 & 1.162 & 0.050 & 0.052 & 1.161 & 0.095 \\
32k & 5.254 & 5.265 & 0.181 & 0.184 & 5.252 & 0.221 \\
128k & 31.755 & 31.774 & 0.484 & 0.492 & 31.715 & 0.605 \\
256k & 92.108 & 92.188 & 0.557 & 0.574 & 92.047 & 0.790 \\
\bottomrule
\end{tabular*}
\end{table*}

Cross-replica reuse removes most of the latency otherwise spent recomputing a long prefix on the receiving replica. At 128k and 256k, TTFT falls to approximately 0.605 and 0.790 seconds. This comparison is a controlled shared-prefix hit against a locally cold receiving replica, not an average over production traffic.

Cold medians remain close across arms. The 8k case retains a startup fluctuation, visible among the individual observations. First local reuse has ON-minus-OFF median differences of approximately 2--17 ms across input sizes. These small differences help describe the observed tradeoff, but the experiment has too few independent blocks to estimate a stable overhead.

\subsection{Session time}\label{session-time}

\begin{figure*}[!t]
\centering
\includegraphics[width=1.0\textwidth]{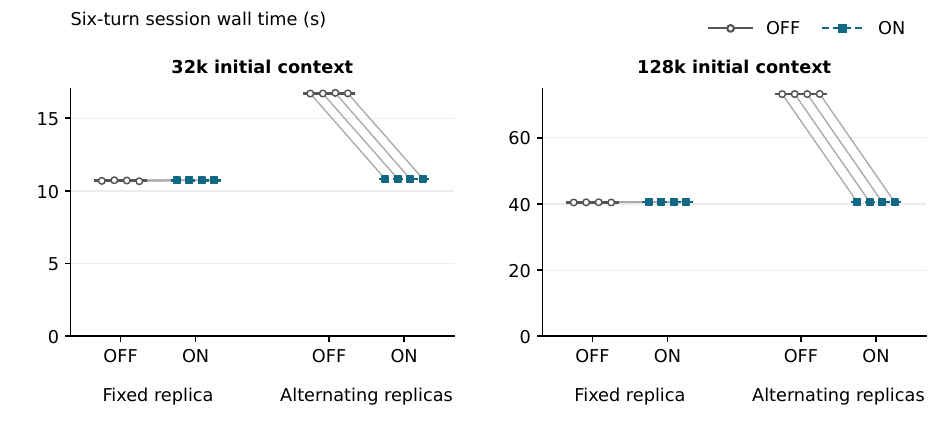}
\caption{Six-turn wall time at two initial context lengths. Thin lines connect each ON/OFF pair within a routing condition; short horizontal marks show medians. Fixed and alternating workloads use independent namespaces and are not cross-route payload pairs.}
\label{fig:4}
\end{figure*}

Figure~\ref{fig:4} and Table~\ref{tab:4} report the six-turn session times.

\begin{table*}[!t]
\centering\small\setlength{\tabcolsep}{5pt}
\caption{Four paired sessions per cell. Negative changes indicate shorter wall time.}
\label{tab:4}
\begin{tabular*}{\textwidth}{@{\extracolsep{\fill}}llrrr@{}}
\toprule
\textbf{Initial context} & \textbf{Placement} & \textbf{OFF median (s)} & \textbf{ON median (s)} & \textbf{Mean paired change} \\
\midrule
32k & Fixed & 10.706 & 10.742 & +0.36\% \\
32k & Alternating & 16.708 & 10.821 & −35.26\% \\
128k & Fixed & 40.465 & 40.605 & +0.34\% \\
128k & Alternating & 73.220 & 40.553 & −44.61\% \\
\bottomrule
\end{tabular*}
\end{table*}

The session-level improvement is smaller than the isolated cross-replica TTFT improvement because the session includes initial cold computation and other work. Analysed separately, the two block pairs yield alternating-session reductions of 35.21--35.30\% at 32k and 44.58--44.64\% at 128k. This is a consistency check within the existing experiment. Fixed placement already preserves local prefix state, leaving little room for the shared path to help.

\subsection{Long output and concurrency}\label{long-output-and-concurrency}

Measured decode-rate changes span approximately −0.82\% to +0.16\%, with two paired observations per cell. All 32 forced-length outputs contain role-like delimiters and literal \paperCode{<think>} strings, and full outputs differ in all 16 ON/OFF phase pairs. Cold/hot outputs match within each arm, including zero-hit cold computation, so the text differences alone do not identify a cache-transfer fault. This workload compares rates at equal token counts; it does not assess normal long-answer quality or output equivalence.

The short concurrency windows show substantially higher output tokens per second under alternating placement: mean paired gains are approximately 87\% with four sessions and 78\% with eight. Fixed-placement window throughput changes by less than 0.4\% in the negative direction. These rates include initial cold prefill and come from only two short windows per condition; they are not steady-state decode capacity or robust tail-latency measurements.

\section{Discussion and Limitations}\label{discussion-and-limitations}

The practical lesson is to validate representation, ordering and full-size allocation separately, then test service outcomes. Registration and cache-hit counters describe availability of a reuse path; byte probes establish whether a particular path preserves state. The controlled ordering intervention strengthens that diagnosis without requiring a new copying algorithm.

The performance result turns on locality. When the local prefix remains available, fixed placement provides nearly all the measured benefit. Shared RAM greatly reduces prefill after a controlled replica switch. This supports combining affinity with a shared reuse path where placement must change. Actual gateway affinity, load-driven migration and cache-pressure revisits remain to be evaluated.

The performance experiment uses one model configuration, one host and two block pairs. The two serving GPUs were isolated from normal requests, but the host was not exclusive: another GPU reached approximately 24\% mean utilization in one suite window. CPU binding and host-memory page placement were not recorded. These conditions especially limit interpretation of fractional-percent differences. Saturation beyond the 256 GiB pool, sustained open-loop capacity and cross-host sharing were not tested; native offload was not a matched third arm.

Correctness evidence has distinct scopes. The copy probes test single byte objects rather than every registered kernel group. No transfer trace of the original failed service request was captured, so the probes do not attribute every historical failure to this fallback. Of the performance run's 736 exact-answer checks, 448 later-turn requests already contain the expected values in assistant history, so those checks do not independently verify distant-prefix recovery. The separate 264-request repair regression retains its own fixtures and counts. The synthetic sessions use canonical histories with thinking disabled; they measure neither model quality nor real coding-agent task completion.

\section{Conclusion}\label{conclusion}

Two replicated 27B inference engines can reuse shared host-memory state effectively after the transfer path is validated. This deployment required an existing representation patch, an allocator choice that restores stream ordering, and a full-capacity pinned-allocation check. Controlled experiments show the largest benefit when a request moves to a replica without its prefix: long-context TTFT and alternating-session time fall substantially, while fixed placement gains little. The engineering result is a validated configuration and a method for distinguishing cache correctness from locality-dependent performance.

\section*{Acknowledgements}

The author acknowledges Research Technology Services, UNSW Sydney, for providing GPU resources on the Katana computational cluster \cite{ref12} for model quantization.

\textbf{AI assistance.} OpenAI Codex assisted with developing and refining the experimental design, implementing and executing benchmark and diagnostic scripts, analysing results, and drafting and revising the manuscript. Anthropic Claude provided feedback on the manuscript. The author takes responsibility for the methods, results, and final text.

\end{document}